\documentclass[12pt,a4paper]{article}
\usepackage{ocg}
\usepackage{times}
\usepackage[english]{babel}   
\usepackage{graphicx}
\usepackage[flushmargin]{footmisc}
\begin{document}
\renewcommand\thanks[1]{\renewcommand\thefootnote{\arabic{footnote}}\footnote{#1}\noindent}
\uppercase{\title{Shopping uncertainties in a mobile and social
context}}
\author{
  Mauro Cherubini, Rodrigo de Oliveira, and Nuria Oliver\footnote{Telefonica Research, via Augusta 177, Barcelona, SPAIN
 \texttt{ \{mauro, roliv, nuriao\}@tid.es}}
}
\maketitle
\addtocounter{footnote}{1}
\thispagestyle{empty}
\bibliographystyle{plain}
\begin{abstract}
We conducted a qualitative user study with $77$ consumers to
investigate what social aspects are relevant when
supporting customers during their shopping activities and particularly in situations when they are
undecided. Twenty-five respondents ($32\%$) reported seeking extra
information on web pages and forums, in addition to asking their peers
for advice (related to the nature of the item to be bought).
Moreover, from the remaining $52$ subjects, only $6$ ($8\%$) were
confident enough to make prompt comparisons between items and an immediate purchasing choice, while
$17$ respondents ($22\%$) expressed the need for being away from persuasive elements. The remaining $29$ respondents ($38\%$) reported having a suboptimal strategy for making their shopping decisions (\emph{i.e.} buying all items, not buying, or choosing randomly). Therefore, 
the majority of our participants ($70\% = 32\% + 38\%$) had social and information needs when making
purchasing decisions. This result motivates the development of applications
that would allow consumers to ask shopping questions to their social network while on-the-go.
\end{abstract}

\section{Introduction}
Shopping can be seen as a social activity. When conducted
with friends or family (henceforth called \textit{social shopping}),
it supports quality time with peers \cite{Kirtland:1996fx} and often
leads to purchases that one would not have been made alone
\cite{Mangleburg:2004kx}.
Furthermore, many shopping activities are often conducted in
physical shops rather than online\footnote{In the third quarter of
2008, the Department of Commerce revealed that U.S. retail
e-commerce sales totaled \$$31.6$ billion, adjusted. E-commerce
sales accounted for $3.1$ percent of total sales
\cite{Winters:2008qa}.}. Being physically in the shop allows buyers
to fully \textit{experience} (\textit{e.g.}, with smell or touch) the product or the service
before committing on buying~\cite{Fano:1998ve}.

We shall highlight two challenges
that prevent large-scale adoption of m-commerce applications: lack of
technological standards \cite{Jones:2008eu} and {\em limited
understanding of a shopper's context}. In this paper, we address the second challenge.
In particular, a recurrent issue when designing applications
for social and ubiquitous activities is related to understanding the
\textbf{moments of need} when users might require support.
O'Hara and Perry \cite{OHara:2003pi} studied the moments when
shopping impulses are deferred, demonstrating how transactions are
often deferred when there is a lack of information. They also
highlighted how shopping is a social and collaborative
activity and how shoppers often deferred purchases because they
wanted to ask their friends and relatives for advice.

The study presented in this paper extends O'Hara and Perry's work: while they asked
subjects to document situations in which they deferred a purchase,
we focus on situations where consumers are physically in the shop
and are \textbf{undecided on what to buy}. Particularly, we are
interested in investigating the \textbf{role that the buyer's social
network plays on how s/he solves this particular moment of impasse}.


%

\section{Methodology of the study}
We conducted a combination of qualitative research methods including
an online questionnaire and interviews \emph{in situ} with a sample
of consumers in Barcelona, Spain. The questionnaire contained
demographic questions (\textit{e.g.} age, gender and occupation of
the respondent), followed by shopping related questions. The two
items that represented the core of the questionnaire are: 1)
\textbf{Q1: What do you do when you are out shopping and you are
undecided between different options?} and 2) \textbf{Q2: How
important are your friends and family members' opinions in
influencing your shopping choices?}
In the introduction of the questionnaire, we emphasized to our respondents
that we were interested in shopping situations that were not repetitive
(\textit{e.g.}, buying grocery items), and where they were not acquiring unique and expensive
items (\textit{e.g.}, buying a car or a house). Additionally, we framed Q1 with a specific task
(\textit{e.g.}, being in a shop with the purpose of buying something and
being undecided), because we wanted to focus only on shopping activities where the buyers'
decision processes would follow rational criteria ({\em e.g.} buying the cheapest,
the most fashionable, the most durable, etc., item with the smallest effort).
We are aware that in many shopping situations buying decisions are taken
regardless of objective standards (see for instance the interpersonal dilemmas
described by Prus \cite{Prus:1993la}).

Seventy-seven people (m: $56$, w: $21$) filled the online
questionnaire. Their median age was $32$ years (min: $22$, max:
$50$) and the occupation of respondents was fairly diverse, ranging
from administrative assistants, designers, engineers, researchers,
managers, students and accountants. From this sample, we selected $13$ subjects 
who would represent the highest variability in occupation, gender, and age. These subjects were further invited to an interview session where we elicited additional information about the participants' shopping behaviors. The interviews were conducted in the shops typically
visited by the interviewees in order to place the conversation in the 
natural shopping location. With this mixed sample and situated context, we
were hoping to address as many shopping-related factors as possible.

\section{Results and Discussion}

\begin{figure}[h]
\begin{center}
\leavevmode
\includegraphics[scale=0.85]{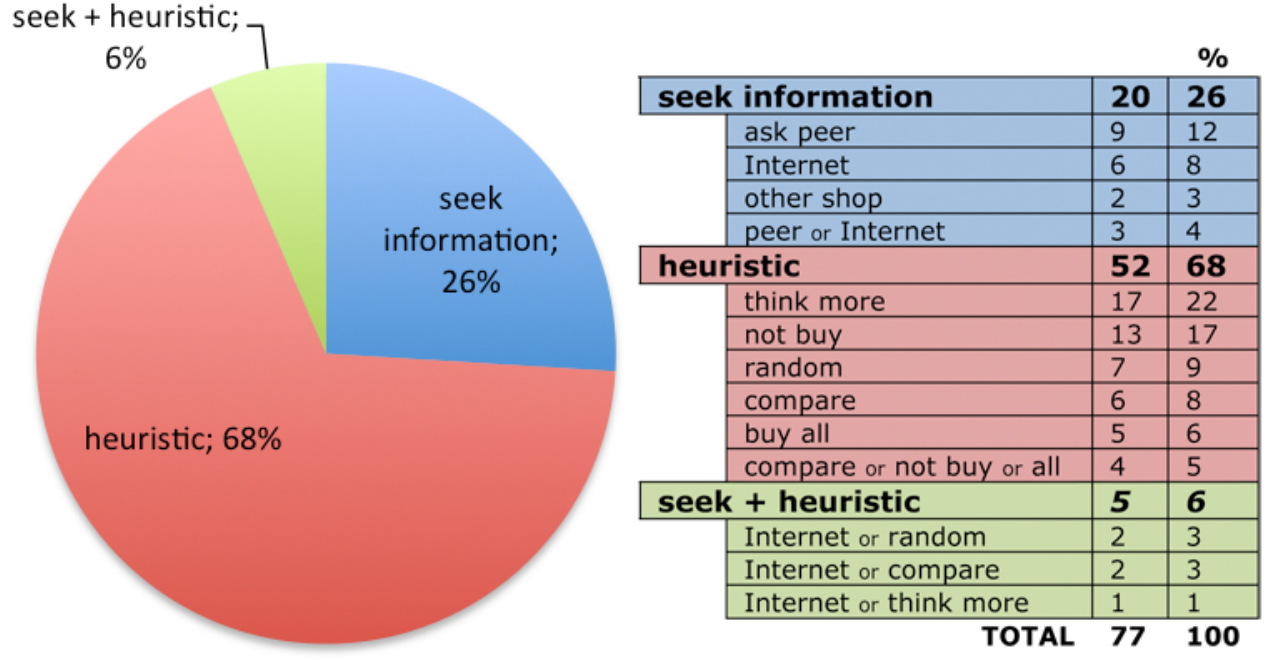}
\end{center}
\caption{Left, pie chart of the responses collected for Q1. Right,
details of the individual responses for each category.}

\label{piechart}
\end{figure}

The answers to the above-mentioned research questions (Q1 and Q2)
were manually categorized by grouping responses with a similar argument --
but eventually different formulation.
Figure \ref{piechart} summarizes the results with respect to \textbf{Q1}. 
In the Figure, we have classified the participants answers into $3$ categories: 
$52$ respondents ($68\%$ of the sample)
reported using one or multiple \textbf{heuristics}\footnote{None of
the heuristics indicated the need to look for extra help or
information.} to solve the moment of impasse (category $1$); $20$ respondents
($26\%$) declared they would \textbf{seek extra information} beyond
what was available at the moment (category $2$); and $5$ respondents ($6\%$) reported 
looking for information in addition to the heuristics (category $3$). 
Moreover, we have identified a number of subcategories in each of these categories.

{\bf Heuristic Category}~~$17$ respondents (out of the $52$) said they would \textit{go home and think some
more}\footnote{It wasn't clear if this thinking involved consulting
extra sources of information or not. We assigned them to the
heuristic category because the answers revealed an inclination
towards expressing the need to be away from persuasive elements.}.
Another group of $17$ respondents ($13 + 4$) reported that when undecided, they
preferred \textit{not to buy}.
Ten respondents ($6 + 4$) reported carefully evaluating the items by
\textit{comparing} the price, the quality of the materials,
etc., and finally choosing the cheapest item or that with the best
price-quality ratio.
Finally, $9$ respondents ($5 + 4$) said they would simply
\textit{buy all} the items, assuming their price was reasonable in
relation to their budget, while $7$ others reported simply buying
one of the items at \textit{random}, without spending too much time
thinking about it. These results indicate that \emph{most participants had
a suboptimal methodology to promptly solve the moment of impasse}
(\emph{i.e.}, choose items randomly, buy all, give up choosing), and
only $10$ felt confident enough to choose among the items by comparing them.

{\bf Seek Additional Information Category}~~Participants in this group invoked two major
sources of information: $12$
participants ($9 + 3$) reported \textit{asking advice to their friends or
family members}. Another $9$ respondents ($6 + 3$) reported \textit{seeking
information on the Internet}. Only 2 respondents said that they
would \textit{browse more shops} in order to find better deals.
Figure \ref{piechart} summarizes the results of Q1.


If we merge those who explicitly reported seeking information and
those who were not confident when comparing items and making a purchase decision 
promptly -- by not buying anything, buying randomly, or buying all, we obtain that a
total of $54$ respondents ($20 + 5 + 13 + 7 + 5 + 4$), corresponding to $70\%$ of the sample,
reported not having an optimal strategy when uncertain on what to buy. This result 
supports the idea that consumers could benefit from a mobile application
that would provide them with extra information while they are shopping.

With regard to \textbf{Q2}, $9$ respondents ($12\%$) reported
trusting their peers for \textit{all their buying activities}. The
majority of respondents ($47$ or $61\%$) reported caring about the
expertise of their friends \textit{only on certain topics}.
Therefore, they selectively asked for advice about specific products
to some of their friends or family members, depending on their
expertise. Thirteen respondents ($17\%$) reported asking some of
their peers for \textit{shopping advice}, but not necessarily in
relation to specific products. Finally, $8$ respondents ($10\%$)
reported \textit{rarely asking for advice} to their friends and family
members. When we combine the results of Q1 and Q2, we observe that
a large majority of those who reported using a heuristic to solve
the shopping impasse also reported asking for advice to their social
network under specific or all shopping circumstances (see Table \ref{crosstab}).
These results reinforce the proposal that consumers might benefit from a
mobile application that would allow them to ask shopping-related questions
to their peers while on-the-go.

\begin{table}[htdp]
\caption{Q1 $\times$ Q2 crosstabulation. Note how for Q1, the
heuristic category has only 7 (out of 52, or 13\%) respondents who
would rarely seek their peers' advice.}
\begin{center}

\begin{tabular}{c}
\includegraphics[scale=0.8]{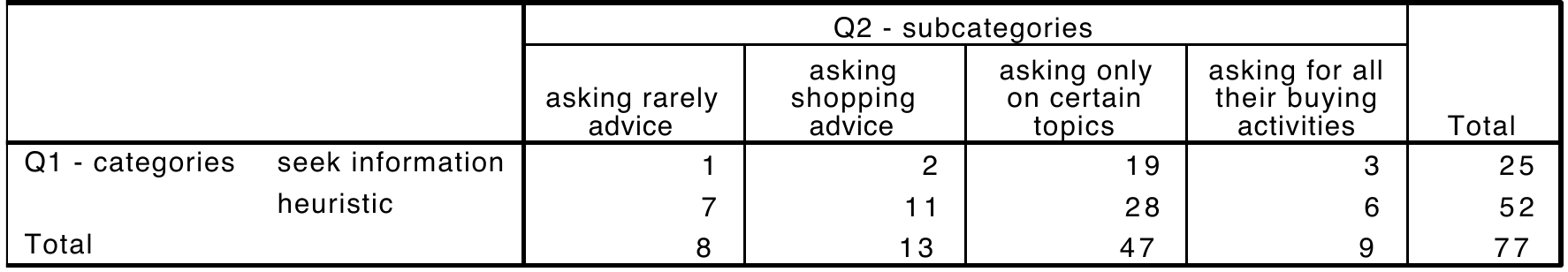}
\end{tabular}
\end{center}
\label{crosstab}
\end{table}%

During the interviews, we had the opportunity to further discuss the
relation between the need for advice from peers and the nature of
the items that participants were interested in buying. Almost all of
the interviewed participants stated that their need for
shopping-related advice very much depended on the kind of items
they intended to buy. Interestingly, few interviewees
reported knowing their taste well and therefore rarely requiring
advice from others when buying clothes. Conversely, they defined
themselves as less experts in other kinds of products
(\textit{e.g.}, technology) and therefore were more inclined to ask
more knowledgeable friends for advice when shopping such items.
During the interviews we also identified two factors that have an
impact when asking for advice: \textit{the price of the item}, and
\textit{the uncertainty related to the specific use of the item} to
be bought (\textit{e.g.}, a present). Additional factors that made
the buyer ask his/her social network for advice included: specific
features of the item, the expertise of the buyer in relation to the
category of the item and the context of its use.

\section{Conclusions and Future Work}
In this paper, we have presented the results of a user study with $77$ consumers
aimed at investigating what social aspects play a role in their shopping decisions.
The user study revealed that consumers often need additional information 
about the products they intend to buy while
shopping on-the-go. While one third of the sample ($25$ out of $77$) reported
the \emph{need for increasing their knowledge of the product before
committing on buying}, another third could potentially follow the
same heuristic. This is consistent with the findings of O'Hara and
Perry \cite{OHara:2003pi}. Additionally, respondents gave priority
to two product-related information channels: (a)
online web pages and forums, and (b) \emph{advice from their friends
or family members.} Although designers have been exploring option
(a) for some time
\cite{Brody:1999yq,Fano:1998ve},
these results indicate that option (b) is a promising area to be
explored.
Aside from the fact that consumers can ask peers for advice through
phone calls and multimedia messaging services, we believe that there
are opportunities for developing multimedia-based mobile
applications for social shopping. This belief is supported by an
interesting and somewhat expected result: the need for advice from
peers is related to \textit{the nature of the item to be bought}.
Therefore, mobile applications for social shopping would benefit
from understanding the link between the kind of item to be bought
and other variables of the shopper's context (\textit{e.g.}, the
trust assigned to a certain peer giving advice on that item).
We are planning to explore more deeply this relation and to validate
the results of this study with a large sample of consumers.




%
%
\bibliographystyle{plain}

\pagebreak

\end{document}